# IMPACT OF E-GOVERNMENT SERVICES ON PRIVATE SECTOR: AN EMPIRICAL ASSESSMENT MODEL

## Hussain Wasly[1] and Dr Ali AlSoufi[2]

[1]Asia e University, Malaysia
[2]University of Bahrain, Bahrain

## ABSTRACT

*Despite the large investments in the field of e-Government (e-Gov) around the world, little is known about the impact such investment. This is due to the lack of guidance evaluation, absence of appropriate tools to measure the impact of e-Gov on the private sector, as well as the lack of effective management to resolve or eliminate the barriers to e-Gov services that led to the failure or delay of many projects. This paper is primarily concerned in determining the impact of e-Gov services on the private sector. A combination of Modified Technology Acceptance Model (TAM), DeLone and McLean's of IS success will be utilized as a research model and e-Gov Economics Project (eGEP) framework to measure "Efficiency, Democracy & Effectiveness impact" for G2B services. The research result will help e-Gov decision makers to recognize the critical factors that are responsible for G2B success, specifically factors they need to pay attention to gain the highest return on their technology investment, hence enabling them to measure the impact for e-Gov on the private sector. The paper has also demonstrated the usefulness of Structural Equation Modeling (SEM) in analysis of small data sets and in exploratory research.*

## 1. INTRODUCTION

The financial and economic crisis beginning in 2008 has forced government and private sector as well to focus on how to maximize saving costs and providing good services. Countries spend millions and even billions on IT and e-Gov programs, for example in 2009; the US government spent more than $71 billion on IT, with an estimated 10 percent of it related to e-Gov which means around 7.1 billion for e-Gov. In 2014, the total IT investment in USA Federal government is $81,996 million with a modest 2.1% increase over fiscal year 2012 [1]. E-Government (e-Gov) refers to the use of information and communication technologies, particularly the Internet, to deliver government information and services [2]. E-Gov can create meaningful and big benefits around the world for governments, businesses, and citizens [3].
Government to business (G2B) impacts many areas like satisfaction / willingness to remain using, time saving / cost reduction, integration with the existing business processes, trust, security, expenditure & labour invested [4]. E-Gov investments could easily be recovered if Governments are able to do impact assessments from first stage and measuring the impact their e-Gov services. According to Chang-hak Choi [5], South Korea invested $80 million to implement e-Procurement; as a result it was able to do savings in 2009 amount of $3.2 billion, which means South Korea recovered the cost in 10 days.

According to European commission [6], many factors affect positive G2B impact and one of them is e-Gov barriers. The barriers to e-Gov project team have identified seven key categories of barriers that can block or constrain progress on e -Gov as the following: "1) Leadership failures 2) Financial inhibitors 3) Digital divides & choices 4) Poor coordination 5) Workplace and organizational inflexibility 6) Lack of trust and 7) Poor technical design". These have been derived from a broad review of the literature and research on e-Government, supplemented by an analysis of the experience and knowledge of the partners in the project, including the reaction of growing stakeholders obtained from the expert group workshops and project work. Furthermore lack of clear or good measurement framework is another factor that affects positive G2B impact. Therefore, many countries have established national measurement Frameworks to identify the benefits and returns of investments of e -Gov services, each one measuring from different angles.

According to Heeks [7], some of the well-known national measurements methodologies are MAREVA (A Method of Analysis and Value Enhancement) developed by the French, Electronic Administration Development Agency (ADAE) and Bearing Point (2005), WiBe Economic Efficiency Assessment methodology (Federal Ministry of the Interior, Germany, 2004), used by the German federal administration, eGEP measurement framework developed by the European Commission [8] on the basis of a review of MAREVA, WiBe, and other frameworks developed in the UK, Holland, and Denmark.

MAREVA measurements methodology is built around return of investment (ROI) which provides a method for agency to compute costs and gains. This method provides a way to calculate the expected return on investment (ROI) before a project is taken up. However, it suggests additional four parameters to measure a project requirement as level of risk, gained benefits to employees and society, and real benefits to clients. Each of the five parameters is rated on a five point scale as a radial diagram for all projects being compared. The key benefits for the clients are identified as saving of time, saving of cost and simplification of accessibility. WiBe is a measurement methodology in Germany for assessment of IT projects. It provides different templates to calculate costs and revenues. These templates are useful to develop the method of assessing investments, operating costs, and revenue impacts for the agency. The eGEP framework is built around the three value drivers of efficiency (organizational value), democracy (political value), and effectiveness (user value), and it is "elaborated in such a way as to produce a multidimensional assessment of the public value potentially generated by e-Government, not limited to just the strictly quantitative financial impact, but also fully including more qualitative impacts." [38].

There have been large investments in the field of IT and e-Gov in all parts of the world. However, little is known about the impact of investments in e-Gov, due to lack of guidance evaluation, and the absence of appropriate measurement tool for the impact of e-Gov on the private sector, as well as the lack of effective management to resolve or eliminate the barriers to e-Gov which led to the failure or delay of many projects, especially in developing countries. Many government projects fail for various reasons. These include unclear business cases, misaligned accountability and motivation structure, management and lack of technical expertise by external service providers, poor discipline of project management, inadequate tracking systems and performance management practices, uncertain budget environments and ineffective governance [9].

According to Valentina [10] study, there is a positive impact and many benefits by using e-Gov services such as cost saving and efficiency gains, quality of service delivery to businesses, citizens and government, transparency, anti-corruption and accountability, increase government capacity, improve decision making quality, creation of networked community and promote use of ICT in other sectors of the society. In terms of cost saving and efficiency gains, there is a new system for Beijing's business e-Park that applies the latest computer and Internet technologies to improve the efficiency and responsiveness of government. By using that system businesses can reduce the time required for gaining approval for specific applications from 2-3 months to few days.

According to European commission [6], there are many factors that affect or hinder the positive G2B impact and one of them is e-Gov barriers. The barriers to e-Gov project team have identified seven key categories of barriers that can hinder or constrain progress on e-Gov. These keys are leadership failures, financial inhibitors, digital divides & choices, poor coordination, inflexible workplace and organizational, lack of trust, and poor technical design.
The objectives of this research are mainly to measure the impact and net benefits of G2B services on private sector by using the proposed model.

## 2. RESEARCH METHODOLOGY AND MEASUREMENT FRAMEWORK

Partial least squares structural equation analysis (SmartPLS Version 2.0.M3) was used as statistical technique for this study to analyse the information gathered from the surveys and interviews. The PLS guidelines prescribed by some researchers [11] [12] were followed. PLS is the better SEM technique when "hypotheses are derived from macro level theory in which all salient and/or relevant variables are not known"; "relationships between theoretical constructs and their manifestations are vague"; and "relationships between constructs are conjectural" [13]. Measurement conditions consider the characteristics of the latent and manifest variables. PLS is best suited when "some or all of the manifest variables are categorical or they represent different levels of measurement"; "manifest variables have some degree of unreliability" and "residuals on manifest and latent variables are correlated" [13]. Falk and Miller [13] theorize one distribution condition in which PLS is better suited: "data come from non-normal or unknown distributions". PLS is also more appropriate when these practical conditions are present: "cross-sectional, survey, secondary data, or quasi-experimental research designs are used"; "a large number of manifest and latent variables are modelled"; and "too many or too few cases are available" [13].

In this research the authors used a combination of the qualitative & quantitative research methods (Mixed Method Approach) to overcome certain disadvantages of each method.

### 2.1. Technology Acceptance Model (TAM)

TAM primary objective is to predict and explain the use of technology [14]. In TAM, perceived usefulness is defined as "the degree to which a person believes that using a particular system would enhance his or her job performance" [15], while perceived ease of use is defined as "the degree to which a person believes that using a particular system would be free of effort" [15]. TAM research has confirmed perceived usefulness as a key and consistent predictor of IT usage intention during the initial and later stages of usage [16]. Behavioural intention to use the system has been studied extensively in the IS literature. Klopping M. and McKinney E.[17] proposed modified TAM for e-Commerce . They made two common modification of the original TAM to fit it with the online shopping domain. To further enhance the model for e-commerce use, they also modify the TAM in an important and unique way. They add a direct effect of perceived usefulness on actual use. Consumers may view online shopping as a necessity even if their intention to use the technology is relatively unchanged. That is, some consumers may report that they do not have an improved intention toward online shopping, while at the same time increasing their actual online shopping use. According to Davis [15] perceived usefulness and perceived ease of use effect the actual outcomes.

## 2.2. DeLone and McLean IS Success Model

Evaluating the success of information systems remains a challenging task for researchers and interested as well. Government and companies invest a lot in information systems to get a desired return on their investment. Numerous studies were conducted to assess IS success [18] [19]. In order to simplify the model, DeLone and McLean grouped customer, societal, inter-organizational, and industry impact into "net benefits". The Service quality was also included in the model based on the importance of service as an important aspect of the success of the information system. The "use" has been divided into intention to use and use components.

## 2.3. E-Government impact Measurement Framework

In many cases no clear or good measurement framework is another factor that affects a positive e-Government impact. Therefore, many countries have a national measurement Frameworks to identify the benefits and returns of investments of e-Gov services, each one measuring from different angles.

The eGEP framework as shown in Figure 2.1 [8] is built around the three value drivers of efficiency (organizational value), democracy (political value), and effectiveness (user value), and it is "elaborated in such a way as to produce a multidimensional assessment of the public value potentially generated by e-Government, not limited to just the strictly quantitative financial impact, but also fully including more qualitative impacts."

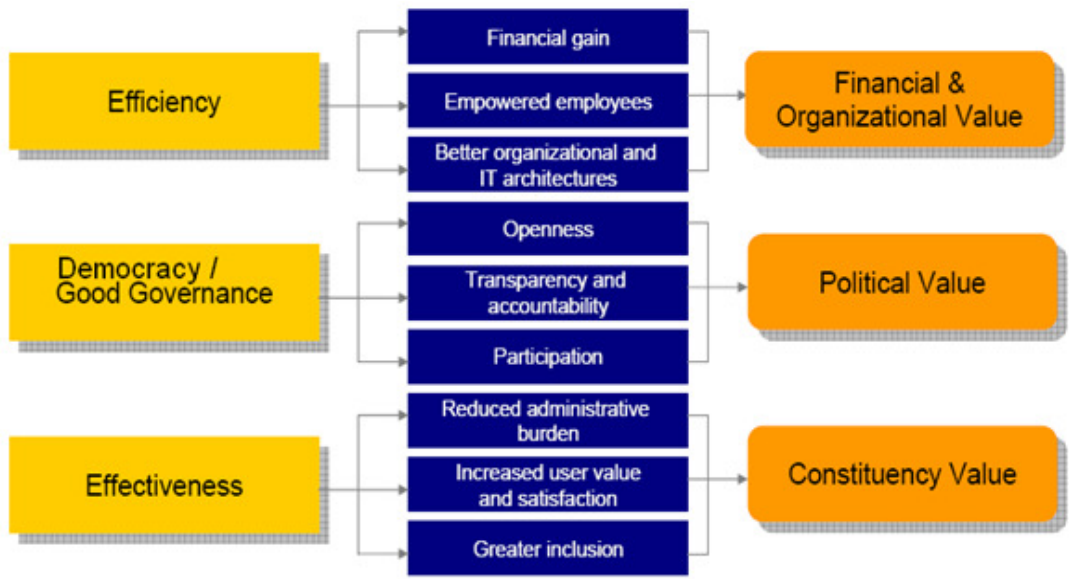

Figure 2.1: Adapted from e-Gov Economics Project. Measurement Framework, Final Version, May 2006.

The eGEP model built around the three value drivers of efficiency, democracy/good governance, and effectiveness and elaborated in such a way as to produce a multidimensional assessment of the public value potentially generated by e-Gov, not limited to just the strictly quantitative financial impact, but also fully including more qualitative impacts [20].

## 2.4. Research Model

In order to provide a general and comprehensive definition of IS success and user acceptance that covers different perspectives of evaluating information systems, moreover in order to create more comprehensive and solid model for evaluating IS success and evaluate the quantitative and qualitative impact on private sector, this research uses an extension of DeLone and McLean's model of IS success, Modified TAM including efficiency, governance / democracy, and effectiveness of eGEP framework including customer stratification. An eGEP framework is the selected tool to measuring net benefits. The resultant (combined model) is shown in Figure 1.

Some of researchers had integrated TAM and IS success models together in their studies. According to Wang and Liu [22], Both TAM and the D&M update IS success model have their own strengths and weaknesses in terms of evaluating the success of an information system because TAM was mainly developed to focus on evaluating system usage from users' perspective, while D&M update IS success model concerns about the relationships among actual system usage, user satisfaction, and their influence on the overall benefits. Wang and Liu [22] proposed a model which is an integration of TAM and the D&M update IS Success Model "to create a more comprehensive and solid model for evaluating IS success model, since these two models are complementary to each other in a certain way". Some researchers conclude that "TAM2 and TAM, D&M (1997, 2003) IS success model [22], and Seddon's [23] IS success sub-model have been used together in studies even though TAM2, also IS success model [18] [24] are the extension of TAM and Seddon's (1997) models respectively and contain all the variables for the former models". According to Zaied A. study [25], the proposed model of integration between TAM and D&M IS Success models with two more success dimensions (Management support and Training) has been validated by an empirical study based on a questionnaire.

The research model has been tested using fifteen hypotheses as shown in Figure 3.1 as a research model to measuring e-Gov success and its impact on private sector.

## 3. DATA ANALYSIS AND RESULTS

This presents the analysis of the gathered survey data in this study and analysis of the instrument also the assessment of the empirical model. Also it presents a descriptive statistics of the user data and the model instrument. The model used in this research comprises of 10 latent variables which cannot be directly measured.

## 3.1. Research Instrument and Sample

A total of 174 questionnaires were collected either by email or interviews. Table 3.1 shows the distribution of research sample according to respondents department. The largest group of respondents is Budget & fiscal operations (including Accounting, warehouse, purchasing, Treasury) which accounts for 80% of the responses followed by Human Resources with 13.22%; and government relationship with 9.20%. This is expected since SADAD and other economic/financial services are the most used applications/systems with more than 66%, followed by some services that related to human resources and government relationship.

Table 3.1: Distribution of Respondents by Departments.

| Department | % | # of Respondents |
|---|---|---|
| Planning and building inspection. | 1.72% | 3 |
| Community Development / Economic Development. | 1.15% | 2 |
| Customer Service. | 1.72% | 3 |
| Facility Management. | 2.30% | 4 |
| Admin. & service department | 5.17% | 9 |
| Health & Safety Department. | 1.15% | 2 |
| Procurement | 3.45% | 6 |
| Recruitment | 4.02% | 7 |
| Government relationship. | 9.20% | 16 |
| Human Resources. | 13.22% | 23 |
| Information Technology. | 5.75% | 10 |
| Owner | 1.72% | 3 |
| Other | 3.45% | 6 |
| **Total** | **100%** | **174** |

## 3.2. Reflective Measurement Model Assessment

Reflective measurement model assessment focuses on its validity and reliability. The reflective measurement models' validity assessment focuses on (1) convergent validity and (2) discriminant validity, whilst reflective measurement models' reliability assessment focuses on (1) Internal consistency reliability and (2) Indicator reliability [11]. PLS assesses the reliability and validity of the measures of theoretical constructs and estimates the relationships among these constructs. The average variance extracted (AVE), composite reliability (CR), and the item loadings of the reflective constructs are shown in Table 3.2.

If each item shows a strong relationship on its theoretical construct then convergent validity is expected [26]. In PLS, Convergent validity can be assessed by examining the average variance extracted (AVE) [27] [28] [11]. It is recommended that the AVE value of at least 0.5 indicates sufficient degree of convergent validity, meaning that the latent variable is able to explain more than half of its indicators' variance [29] [11].All AVE above 0.5 which indicates significant degrees.

Discriminant validity is expected when the items show a weak relationship with all other constructs except the one it is theoretically associated. Discriminant validity can be assessed by; (1) the Fornell–Larcker criterion and (2) cross loadings [11] [27]. The Fornell–Larcker criterion postulates that "a latent construct shares more variance with its assigned indicators than with another latent variable in the structural model" [28]. In statistical terms, "the Average Variance Extracted (AVE) of the latent constructs is greater than the square of the correlations among the latent constructs" [27]. The cross loadings refer to the indicator's loadings with its associated latent constructs should be higher than its loading with other remaining constructs. AVE, created by Fornell and Larcker [28], attempts to measure the amount of variance that a latent variable component captures from its indicators relative to the amount due to measurement error [30]. It is recommended that the AVE should be greater than 0.50 which means 50% or more variance of the indicators should be accounted. In addition, "the AVEs of the latent variable should be higher than any correlation among any pair of latent construct" [27].

$$AVE = \Sigma\lambda i^2 / \Sigma\lambda i^2 + \Sigma ivar(\varepsilon i).$$

Table 3.2: **AVE:** Average Variance Extracted, **CR:** Composite Reliability, R²: R Square.

| | Loading | Indicator Reliability (Loadings²) |
|---|---|---|
| Systems Quality (AVE=0.73, CR=0.89) | | |
| SQ1 | 0.91 | 0.83 |
| SQ2 | 0.73 | 0.53 |
| SQ3 | 0.91 | 0.84 |
| Information Quality (AVE=0.83, CR=0.94) | | |
| IQ1 | 0.92 | 0.85 |
| IQ2 | 0.97 | 0.93 |
| IQ3 | 0.85 | 0.72 |
| Perceived Ease of Use (AVE=0.70, CR=0.88) | | |
| PE1 | 0.86 | 0.74 |
| PE2 | 0.86 | 0.74 |
| PE3 | 0.80 | 0.64 |
| Behavioural Intention to Use (AVE=0.86, CR=0.96) | | |
| IU1 | 0.95 | 0.91 |
| IU2 | 0.96 | 0.92 |
| IU3 | 0.92 | 0.85 |
| IU4 | 0.87 | 0.75 |
| Perceived Usefulness (AVE=0.78, CR=0.91) | | |
| PU1 | 0.87 | 0.76 |
| PU2 | 0.89 | 0.8 |
| PU3 | 0.87 | 0.77 |
| Actual Use (AVE=0.84, CR=0.91) | | |
| AU1 | 0.91 | 0.84 |
| AU2 | 0.92 | 0.85 |
| Effectiveness Impact (AVE=0.73, CR=0.91) | | |
| EVAB | 0.58 | 0.34 |
| EVPS | 0.93 | 0.86 |
| EVUS1 | 0.94 | 0.89 |
| EVUS2 | 0.90 | 0.82 |
| EVUS3 | 0.58 | 0.34 |
| Efficiency Impact (AVE=0.65, CR=0.93) | | |
| EFEE | 0.75 | 0.56 |
| EFFG1 | 0.85 | 0.72 |
| EFFG2 | 0.80 | 0.63 |
| EFFG3 | 0.88 | 0.78 |
| EFFG4 | 0.85 | 0.73 |
| EFOT1 | 0.77 | 0.59 |
| EFOT2 | 0.76 | 0.57 |
| Democracy Impact (AVE=0.66, CR=0.85) | | |
| DEOP | 0.84 | 0.71 |
| DEPR | 0.79 | 0.62 |
| DETA | 0.80 | 0.64 |
| User satisfaction (AVE=0.77, CR=0.93) | | |
| US1 | 0.90 | 0.82 |
| US2 | 0.83 | 0.69 |
| US3 | 0.87 | 0.76 |
| US4 | 0.89 | 0.8 |

Discriminant validity is also assessed by compare the calculated AVE with the square of the correlations among constructs. According to Fornell and Larcker [28] criterion, "the AVE of each latent construct should be higher than the construct's highest squared correlation with any other latent construct". Table 3.5 below shows the result of the square root of the AVE given in the diagonals which is higher than the correlation among the constructs. This result indicates further strength of discriminant validity presence. Moreover it validate that the constructs met the criteria for acceptable discriminant validity.

Table 3.3: Correlation among Construct Scores (AVE Extracted in Diagonals)

| | IQ | SQ | PU | PEU | BIU | AU | EI | DI | EVI | US |
|---|---|---|---|---|---|---|---|---|---|---|
| IQ | 0.9135 | | | | | | | | | |
| SQ | 0.3434 | 0.8566 | | | | | | | | |
| PU | 0.6518 | 0.5773 | 0.881 | | | | | | | |
| PEU | 0.316 | 0.4114 | 0.5291 | 0.8411 | | | | | | |
| BIU | 0.2252 | 0.4071 | 0.4864 | 0.8122 | 0.9257 | | | | | |
| AU | 0.3823 | 0.5331 | 0.6757 | 0.559 | 0.5264 | 0.9174 | | | | |
| EI | 0.4225 | 0.6902 | 0.7072 | 0.5247 | 0.5007 | 0.7048 | 0.8096 | | | |
| DI | 0.3423 | 0.7956 | 0.7344 | 0.5271 | 0.4944 | 0.6337 | 0.7781 | 0.8112 | | |
| EVI | 0.3263 | 0.5482 | 0.5334 | 0.5202 | 0.4968 | 0.5192 | 0.754 | 0.6081 | 0.8529 | |
| US | 0.3838 | 0.6176 | 0.6012 | 0.5313 | 0.5345 | 0.5878 | 0.7548 | 0.6309 | 0.831 | 0.8755 |

Discriminate validity is further assessed by comparing Indicator's loadings and its cross loadings. A bootstrap resampling (5000 resamples) was used throughout the study to find out if the indicator's loadings should be higher than all of its cross loadings. To get acceptable standard error estimates, Chin [27] suggests 200 resamples. Whereas Hair, Ringle, & Sarstedt [11] suggest the minimum number of bootstrap samples is 5,000. The correlation matrix highlights the loading of the measurement items on the constructs to which they are assigned in the confirmatory factor analysis. The results suggest that most of indicators loaded higher with its respective latent variable.

Construct reliability was assessed by Composite Reliability (CR) to get internal consistency reliability. All CR values are above the suggested 0.60 for all constructs which suggests that the instrument is reliable when conducting exploratory studies [11] [31].

Construct reliability is further assessed by indicator reliability. The recommended value for indicator reliability is more than 0.6 for an exploratory research [32] [11] [33]. Also, for an exploratory research higher than or equal 0.4 is acceptable [34]. Table 4.8 shows that all indicators except (EVAP) have individual indicator reliability values that are larger than the minimum acceptable level of 0.4 and most of them are more than the preferred level of 0.7.
Overall, there is a significant confidence of the survey instrument quality based on the reliability and validity analyses.

## 3.2. Structural Model Assessment

The structural assessment focuses on (1) R² measure and (2) path coefficients' significance as primary evaluation criteria. Also it focus on (3) predictive relevance [11]. R² measures and the level and significance of the path coefficients are the primary evaluation criteria for the structural model because "the prediction-oriented PLS-SEM approach goal is to explain the endogenous latent variables' variance, the key target constructs' level of R² should be high" [11]. R² results of 0.20 are considered high in disciplines, path coefficients with standardized values above 0.2, and path t-value is above 1.96 for significance level 5 percent are usually [11] [12].

The explanatory power of the structural model is evaluated by examining the squared multiple correlation ($R^2$) value in the final dependent constructs. The $R^2$ measures the percentage of variation that is explained by the model. The $R^2$ for the overall model is 0.3841. Figure 3.1 shows the path coefficients and inside the blue balls the $R^2$.

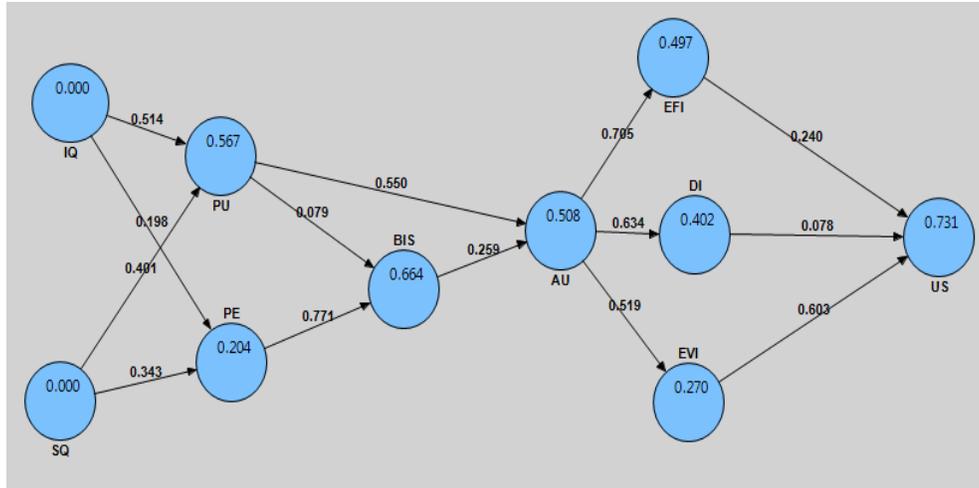

Figure 3.1: Structural Model – Full Model

Table 3.4 below is used to test the research hypotheses and shows results that indicate all reflective items had a significance level greater than .01 and t-values above 1.96 except Democracy Impact-> User satisfaction. In addition, table 3.5 explains which hypotheses were supported.

Table 3.4: Statistical Significance of the Coefficients

| Endogenous Variables | R² | Independent Variables | Standardized Path Coefficients | T Statistics (Inner Model) |
|---|---|---|---|---|
| Perceived Usefulness | 0.57 | Information Quality | 0.51 | 7.6097 |
| | | System Quality | 0.40 | 6.1244 |
| Behavioural Intention to Use | 0.66 | Perceived Usefulness | 0.08 | 1.2148 |
| | | Perceived Ease of Use | 0.77 | 14.9782 |
| Perceived Ease of Use | 0.203 | System Quality | 0.34 | 4.5774 |
| | | Information Quality | 0.2 | 2.5074 |
| Actual Use | 0.51 | Perceived Usefulness | 0.55 | 6.9029 |
| | | Behavioural Intention to Use | 0.26 | 2.5117 |
| User satisfaction | 0.73 | Democracy Impact | 0.08 | 1.0898 |
| | | Effectiveness Impact | 0.6 | 8.1573 |
| | | Efficiency Impact | 0.24 | 2.4763 |
| Democracy Impact | 0.40 | Actual Use | 0.63 | 12.2653 |
| Effectiveness Impact | 0.27 | Actual Use | 0.52 | 8.5546 |
| Efficiency Impact | 0.50 | Actual Use | 0.70 | 11.7332 |

Table 3.5: Research Hypotheses

| Hypotheses | Result |
|---|---|
| H1:  The proposed model is statistically significant. | Supported |
| H2a: Information quality is positively related to perceived usefulness. | Supported |
| H2b: Information quality is positively related to perceived ease of use. | Supported |
| H2c: System quality is positively related to perceived usefulness. | Supported |
| H2d: System quality is positively related to perceived ease of use. | Supported |
| H3a: Perceived usefulness is positively related to behavioural intention to use. | Rejected |
| H3b: Perceived usefulness is positively related to actual usage. | Supported |
| H3c: Perceived ease of use is positively related to behavioural intention to use. | Supported |
| H4: behavioural intention to use is positively related to Actual Usage. | Supported |
| H5a: Actual Usage is positively related to Efficiency Impact. | Supported |
| H5b: Actual Usage is positively related to Governance/Democracy impact. | Supported |
| H5c: Actual Usage is positively related to Effectiveness Impact. | Supported |
| H6a: Efficiency Impact is positively related to User Satisfaction | Supported |
| H6b: Governance/Democracy impact is positively related to User Satisfaction. | Rejected |
| H6c: Effectiveness Impact is positively related to User Satisfaction. | Supported |

The model's capability to predict is another assessment of the structural model. Stone-Geisser $Q^2$ was used to assess the predictive significance of the exogenous variables [36] [37], which postulate that "the model must be able to adequately predict each endogenous latent construct's indicators". Blindfolding is the recommended technique for assessing $Q^2$, The omission distance (D) parameter in PLS should range from 5 to 10 [37]. In this study an omission distance of 10 to run the blindfolding procedure. Table 3.6 shows all $Q^2$ values are greater than zero indicating sufficient predictive power of the structural model exists [12].

Table 3.6: Construct Cross-validated Redundancy.

| Total | SSO | SSE | 1-SSE/SSO |
|---|---|---|---|
| Perceived Usefulness | 522 | 313.2085 | 0.4 |
| Perceived Ease of Use | 522 | 456.1628 | 0.1261 |
| Behavioural Intention to Use | 696 | 338.2507 | 0.514 |
| Actual Use | 348 | 213.9581 | 0.3852 |
| Efficiency Impact | 1218 | 872.8971 | 0.2833 |
| Democracy Impact | 522 | 385.7186 | 0.2611 |
| Effectiveness Impact | 696 | 578.9962 | 0.1681 |
| User satisfaction | 696 | 327.8733 | 0.5289 |

Overall, the reliability and validity analyses demonstrate that there is significant confidence in the quality of the survey instrument.

## 4. Conclusions

The proposed model offers the private sector stakeholders and e-Gov program stakeholders a useful information to determine which factors are important order to gain the highest return and cost saving on their technology investment while ensuring that there is a real impact on private sector . This model shows that e-Gov program holders and private sector should work together in order to get a highest benefits of implementing and using e-Gov services based on a best practices. Moreover, it shows that e-Gov program holders should have a correct measurement model for all e-Gov services; otherwise they will get several issues and private sector resistance.

The paper has also demonstrated the usefulness of Structural Equation Modeling (SEM) in analysis of small data sets and in exploratory research. The author hope that the methodology employed here will provide a useful guide for similar data sets requiring analysis. PLS-SEM is a recommended method when a theory is under development such as the study conducted here. A number of findings related to impact of e-Government on private sector in this study. The findings are discussed below that shows nine of the hypotheses were fully supported and two are not supported.

Hypothesis 1 (H1: The proposed model is statistically significant) measured whether the proposed model is statistically significant. This hypothesis was supported and its $R^2$ is 0.3841 which is above of required value 0.2. $R^2$ results of 0.20 are considered high in disciplines and Path coefficients with standardized values above 0.2 are usually significant and those with values below 0.1 are usually not significant. Values between 0.1 and 0.2 require significance testing [11] [12].

All hypothesis except (H3a) and (H6b) were supported with T-Values above of 1.96.

Hypothesis 2 (H2a: Information quality is positively related to perceived usefulness) and hypothesis 3 (H2b: Information quality is positively related to perceived ease of use) were supported based on the following results:

The hypothesized path relationship between "Information Quality" and "Perceived Usefulness" is statistically significant. The standardized path coefficient for hypothesis (H2a) is (0.51) and its t-value is (7.61) which indicates high significant. Same thing, the hypothesized path relationship between "Information Quality" and "Perceived Ease of Use" is statistically significant. The standardized path coefficient for hypothesis (H2b) is (0.2) and its t-value is (2.51) that indicate a high significant.

Hypothesis 4 (H2c: System quality is positively related to perceived usefulness) and hypothesis 5 (H2d: System quality is positively related to perceived ease of use) were supported based on the following results:

The hypothesized path relationship between "System Quality" and "Perceived Usefulness" is statistically significant. The standardized path coefficient for hypothesis (H2c) is (0.40) and its t-value is (6.12) that indicate a high significant. The hypothesized path relationship between "System Quality" and "Perceived Ease of Use" is statistically significant. The standardized path coefficient for hypothesis (H2d) is (0.34) and its t-value is (4.56) that indicate a high significant.

Hypothesis 6 (H3a: Perceived usefulness is positively related to behavioral intention to use) was not supported. The hypothesized path relationship between "Perceived Usefulness" and "Behavioral Intention to Use" is not significant statistically. The standardized path coefficient for hypothesis (H3a) is (0.08) and its t-value is (1.21) that are below than required standardized coefficient (0.1) and t-value (1.96).

Hypothesis 7 (H3b: Perceived usefulness is positively related to actual usage) was supported based on the following results:

The hypothesized path relationship between "Perceived Usefulness" and "Behavioral Intention to Use" is not significant statistically. The standardized path coefficient for hypothesis (H3b) is (0.08) and its t-value is (1.21) that are below than the required standardized path coefficient (0.1) and t-value (1.96).

Hypothesis 8 (H3c: Perceived ease of use is positively related to behavioral intention to use) was supported. The hypothesized path relationship between "Perceived Ease of Uses" and "Behavioral Intention to Use" is statistically significant. The standardized path coefficient for hypothesis (H3c) is (0.77) and its t-value is (14.98) that are above the required standardized path coefficient (0.1) and t-value (1.96).

Hypothesis 9 (H4: behavioral intention to use is positively related to Actual Usage) were supported based on the following results:
The hypothesized path relationship between "Behavioral Intention to Use" and "Actual Usage" is statistically significant. The standardized path coefficient for hypothesis (H4) is (0.26) and its t-value is (2.51) that indicate a high significant.

Hypothesis 10 (H5a: Actual Usage is positively related to Efficiency Impact), hypothesis 11 (H5b: Actual Usage is positively related to Governance/Democracy impact) and hypothesis 12 (H5c: Actual Usage is positively related to Effectiveness Impact) were supported based on the following results:

The hypothesized path relationship between "Actual Usage" and "Efficiency Impact" is statistically significant. The standardized path coefficient for hypothesis (H5a) is (0.70) and its t-value is (11.73). As well as, the hypothesized path relationship between "Actual Usage" and "Governance/Democracy Impact" is statistically significant. The standardized path coefficient for hypothesis (H5b) is (0.63) and its t-value is (12.26) that indicate a high significant. Moreover, the hypothesized path relationship between "Actual Usage" and "Effectiveness Impact" is statistically significant. The standardized path coefficient for hypothesis (H5c) is (0.52) and its t-value is (8.55) that indicate a high significant.

Hypothesis 13 (H6a: Efficiency Impact is positively related to User Satisfaction), hypothesis 14 (H6b: Governance/Democracy impact is positively related to User Satisfaction) and hypothesis 15 (H6c: Effectiveness Impact is positively related to User Satisfaction) have the following results:
The hypothesized path relationship between "Efficiency Impact" and "User Satisfaction" is statistically significant. The standardized path coefficient for hypothesis (H6a) is (0.24) and its t-value is (2.48). However, the hypothesized path relationship between "Governance/Democracy Impact" and "User Satisfaction" is not statistically significant. The standardized path coefficient for hypothesis (H6b) is (0.08) and its t-value is (1.09) that indicate a high significant. Moreover, the hypothesized path relationship between "Effectiveness Impact" and "User Satisfaction" is statistically significant. The standardized path coefficient for hypothesis (H6c) is (0.6) and its t-value is (8.16) that indicate a high significant.

According to above findings, the "Perceived Usefulness" was strongly affected by "Information Quality" with standardization coefficient of 0.51, followed by "System Quality" with standardization coefficient that equals 0.40. The "Perceived Ease of Use" strongly affected by "System Quality" (Standardization coefficient = 0.34), followed by "Information Quality" (Standardization Coefficient = 0.2). As well as, behavioral intention to use affected strongly by "Perceived Ease of Use", about 0.77 standardization coefficient, and weakly affected or not predicted directly by "Perceived Usefulness" with standardization coefficient of 0.08. Moreover, "Perceived Usefulness" and "Behavioral Intention to Use" are affecting "Actual Usage" strongly, in which "Perceived Usefulness" (standardization coefficient = 0.55) has the strongest effect, followed by "Behavioral Intention to Use" (standardization coefficient = 0.26). The "Actual usage" is affecting "Efficiency Impact" (EFFI), "Democracy Impact" (DI), and "Effectiveness impact" (EFVI) strongly with standardization coefficient 0.70 for (EFFI), 0.63 for (DI), and 0.52 for (EFVI). Finally, "User Satisfaction" affected strongly by "effectiveness Impact" with standardization coefficient 0.6, and then by "Efficiency Impact" with standardization coefficient 0.24 while "Democracy Impact" have weak effect or doesn't predict "User Satisfaction" directly with standardization coefficient 0.08.

Figure 4.1 shows that the revised model based on hypothesizes test results. It shows weak links between "Perceived Usefulness" and "Actual Usage", also between "Democracy Impact" and "User Satisfaction" in this study. However, it shows strong links between other variables.

As future works, this study could be expanded in terms of number of respondents to include different e-Gov programs in different countries. Furthermore the study could go for a better understanding on other segments of the IS business systems out of e-Gov programs to figure out quantitative and qualitative impact, and user satisfaction as well. Moreover, apply research model for government to citizens services with different survey questions that targeting citizens.

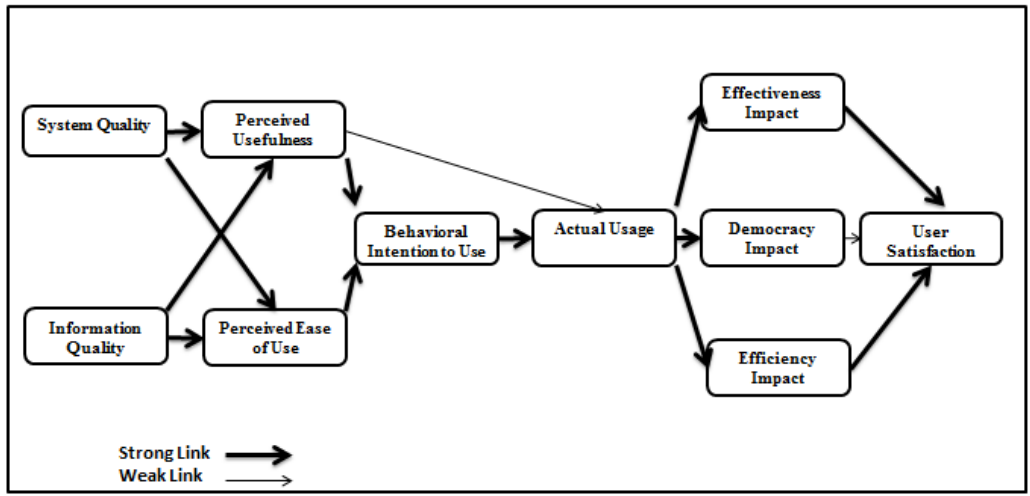

Figure 4.1: Revised Model.

## Authors

Hussain Wasly is a PhD student at Asia e University. He is working for SADAR chemical company in IT department. He worked also for different companies such as Saudi ARAMCO, Marafiq, and ABB where he managed numbers of mega IT/IS Application projects. He leads different teams within organizations. His specializations are enterprise resource management such AS SAP, IT Architecture, IT Project Management, Integrations Systems, and IT Planning. 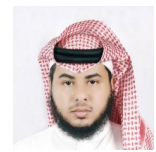

Ali AlSoufi is an assistant professor at University of Bahrain. He has earned his PhD in computer science in 1994 from Nottingham University, UK. He has worked for Bahrain Telecom Co for 8 years as a Senior Manager Application Programme where he overlooked number of mega IS Application projects worked at Arab Open University as the head of IT program and Assistant Director for Business Development during 2007-2010, while working as a consultant for Bahrain e-Government Authority (EGA) in the area of Enterprise Architecture. He is also an active member of the Bahrain National ICT Governance Committee. His specializations is Strategic IT Planning and Governance, IT project management, Enterprise Architecture and Information Systems in Organization. 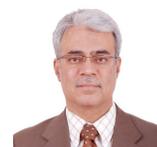